\documentstyle[twocolumn,aps] {revtex}
\begin{document}
\title {Dynamics of Tunneling Centers in Metallic Systems\thanks{
Preprint number : imsc 94/36}}
\author{Tabish Qureshi\cite{email}}
\address{ The Institute of Mathematical Sciences,
 C.I.T. Campus, Madras - 600113, INDIA.}
 \maketitle
\begin{abstract}
        Dynamics of  tunneling  centers  (TC)  in  metallic  systems  is
studied, using the technique of bosonization. The interaction of the  TC
with the conduction electrons  of  the  metal  involves  two  processes,
namely, the screening of the TC by electrons, and the so-called electron
assisted tunneling. The presence  of  the  latter  process  leads  to  a
different form of the renormalized tunneling frequency of  the  TC,  and
the tunneling motion is damped with a temperature  dependent  relaxation
rate. As the temperature is lowered, the relaxation rate per temperature
shows a steep rise as  opposed  to  that  in  the  absence  of  electron
assisted process. It is expected that this behavior should  be  observed
at very low temperatures in a careful experiment. The present work  thus
tries to go beyond  the  existing  work  on  the  {\it  dynamics}  of  a
two-level system in metals, by treating the electron assisted process.
\end{abstract}
\draft
\begin{narrowtext}
\section{Introduction}
        The existence of double-well structures in metals has  now  been
known for a long time. They arise in diverse  physical  situations,  the
most well-known of which is metallic glasses. The existence of tunneling
states in glasses was proposed long back \cite{Black}.  It  is  believed
that  because  the  atoms  in  a  glass  are  `quenched'  in  a   random
configuration, there is a possibility of an atom, or group of  atoms  to
see a complicated potential landscape in its  neighbourhood,  which  may
have at least two neighbouring local minima. In such  a  situation,  the
atom, or the group of atoms can spontaneously tunnel from  one  well  to
the other. In this, and similar situations, the entity is referred to as
a tunneling center (TC). Another well known, and well studied example is
that of hydrogen trapped  in  $Nb(OH)_{x}$.  At  very  low  temperatures
($\sim   70   K$),   the   hydrogen,   which   can   normally    diffuse
quantum-mechanically, is localized in some trap-sites formed  by  oxygen
atoms \cite{Wipf}. The geometry of the system is such that the  hydrogen
can stay in two close-by,  energetically  equivalent  sites  around  the
trapping atom. This results in the formation of a TC. Incoherent neutron
scattering reveals the tunneling of the hydrogen, which is influenced by
the conduction electrons of  the  metal.  These  experiments  have  been
theoretically analyzed using the model of a particle in  a  double-well,
in contact with free electrons  \cite{Grabert,Weiss,tqniobium}.  Lately,
the  conductance  behavior  of  narrow  metal  constrictions  has   been
explained within the framework of  scattering  of  conduction  electrons
from a TC \cite{Ralph}. The behavior of certain heavily doped conducting
polymers  has  also  been   attributed   to   the   existence   of   TCs
\cite{Ishiguro}. Universal conductance fluctuations have also been  used
to  study  the  dynamics  of  a  single  two-level  defect  in  a  metal
\cite{Golding}. In short, tunneling centers in metals have proved to  be
of considerable experimental and theoretical importance.

        Theoretical scenario is the following. The TC can be  classified
to be either {\it  slow}  or  {\it  fast},  depending  on  its  detailed
behavior. When the TC is slow, the effect of the conduction electrons is
to provide some  kind  of  friction  to  the  tunneling  motion  of  the
particle. What actually happens is that the particle being charged,  has
a screening conduction electron cloud around it. As it tunnels from  one
site  to  the  other,  the  electrons  have   to   follow   its   motion
adiabatically. Now it turns  out  that  the  electrons  cannot  actually
follow the motion  of  the  particle  instantaneously  -  the  adiabatic
approximation breaks  down.  The  tunneling  of  the  atom  is  a  local
perturbation which  can  give  rise  to  an  {\it  infinite}  number  of
electron-hole pairs  in  the  electron  gas,  with  {\it  infinitisimal}
energy, and thus to the so-called {\it infra-red divergence}. The result
is that the tunneling of the particle is governed by the  relaxation  of
the conduction electron cloud around it. The tunneling is  thus  heavily
suppressed and at zero temperature the particle tends to localize  in  a
particular well. This small energy behavior of the electron gas  can  be
described in terms of charge density excitations of  the  electron  gas,
which  are  approximately  bosonic  in  nature.  In  addition,  if   the
low-temperature  dynamics  of  the   particle   is   confined   to   the
two-dimensional Hilbert space spanned by the two ground  states  of  the
{\it isolated} wells,  one  can  map  the  problem  onto  the  model  of
dissipative two-level system (TLS) studied extensively by  Leggett  {\it
et al} \cite{Leggett}. One has  to  choose  what  is  called  the  Ohmic
dissipation in order to correctly describe the  behavior  of  electronic
excitations.

        In a slow TC, the atom remains  stationary  during  an  electron
scattering. On the other hand, if the TC  is  {\it  fast},  the  process
where there is a scattering  of  an  electron  simultaneously  with  the
tunneling of the particle, has to be taken into  account.  This  process
was originally proposed by Kondo \cite{Kondo}  and  further  work  along
this  direction  has  been  done   by   Zawadowsky   and   collaborators
\cite{Zawa,Muramatsu,Vladar,Zimanyi,Zarand}. Here the  electrons  assist
the tunneling motion of the particle, and hence the particle  cannot  be
localized in one of the two wells, even at very  low  temperatures.  The
particle, tunneling from one well to the other, scatters  the  electrons
between different orbital angular momentum states. If  we  describe  the
TLS, formed by the particle in a double-well, by a pseudo-spin 1/2,  the
situation is quite similar to  the  Kondo  problem  where  the  impurity
spin-flip causes a spin-flip of the electrons.  Here  the  role  of  the
electron-spin in the Kondo problem is played by  the  different  orbital
angular momentum states of the electrons. In the  present  problem,  the
role of the electron spin itself is rather passive,  and  the  two  spin
states act like two independent {\it channels} of electrons to which the
TC couples. It has been shown earlier that this problem can be mapped on
to  that  of  the  two-channel  Kondo  model,  originally  proposed   by
Nozi\`{e}res and Blandin \cite{Nozieres}, which has come under a lot  of
attention because of its non-Fermi-liquid low temperature behavior.

        In an actual TC in metals, both the processes should be  present
- infact, the amplitude of screening by  conduction  electrons  is  much
larger than that for electron assisted tunneling (because the  tunneling
particle is heavy). The interplay between these two processes may  yield
interesting dynamics at low temperatures. The dynamics of  a  TLS,  with
electron screening, has been studied in detail, and the high as well  as
low temperature behavior is well understood. At very  low  temperatures,
the TLS undergoes weakly-damped coherent  oscillations,  with  a  tunnel
frequency  which  is  strongly  renormalized  by  the  coupling  to  the
electrons. The  damping  coefficient  increases  roughly  linearly  with
temperature, whereas the  effective  tunnel  frequency  follows  a  weak
power-law. As  the  temperature  is  increased,  the  thermally  excited
electron-hole pairs destroy the coherence of the particle. As a  result,
tunneling gets incoherent, and resembles stochastic jump diffusion.  The
jump rate, from one site to the other, however, has a  very  non-trivial
temperature dependence. It follows  a  power-law,  the  jump  rate  {\it
decreasing} with increasing $T$.

        The dynamics of a slow TLS, in all temperature regimes, has been
studied within the so-called dilute bounce gas approximation (DBGA),  to
the underlying functional integral expression for the  dynamics  of  the
TLS  \cite{Leggett,Grabert}.  For  weak  coupling  of  the  TLS  to  the
electrons, some calculations have gone beyond the  DBGA,  to  study  the
low-temperature dynamics \cite{Weiss,pramana}. These two  approaches  do
not  treat  the  electron  assisted  tunneling  process.  Very  recently
conformal field theory has been used to study the very  low  temperature
scaling behavior of  the  two-channel  Kondo  model  \cite{Ludwig}.  The
results obtained are relevant for the anomalous resistivity of the metal
arising out of the  scattering of electrons off the TLS.  This  approach
has been highly successful. We are, however, interested  in  the  actual
dynamics of a TLS and not on its effect on  the  metallic  system.  Such
interest stems from the  fact  that  while  experimentally  probing  the
dynamics of a TLS,  one  usually  deals  with  things  like  the  tunnel
frequency and the relaxation rate of the tunneling atom.

        In this paper, we calculate the Laplace transform  of  the  time
correlation function of the  TLS,  which,  in  the  context  of  neutron
scattering from the tunneling particle, is  related  to  the  incoherent
structure factor. The  correlation  function  will  give  the  necessary
information regarding the interplay between the tunneling  dynamics  and
the relaxation behavior, so commonly seen  in  all  quantum  dissipative
systems.

\section{The Model}

        In the following we assume that the  TC  consists  of  a  single
particle in an asymmetric double well potential. The temperature is  low
enough for the dynamics of the particle to be effectively  described  by
the Hilbert space spanned by the two lowest states of the {\it isolated}
wells. In this approximation, the TC becomes a TLS. Now, the interaction
of the TLS with  the  conduction  electrons  can  be  described  by  the
following Hamiltonian (see e.g., \cite{Muramatsu})

\begin{eqnarray}
H'&=&{1\over 2}\hbar\epsilon\sigma_{z} + {1\over 2}\hbar\Delta_{0}\sigma_{x}
\nonumber\\
&&+ {1\over 2}{\sqrt{2K}\over\rho}\sigma_{z}\sum_{p,k,k'}
(c_{pk'+}^{\dagger}c_{pk+} - c_{pk'-}^{\dagger}c_{pk-})\nonumber\\
&& +{1\over 2}{\hbar\Delta_{1}\over\rho\epsilon_{F}}\sum_{p,k,k'}(\sigma_{+}
c_{pk'-}^{\dagger}c_{pk+}^{\ } + \sigma_{-}c^\dagger_{pk'+}c_{pk-})\nonumber\\
&& +\sum_{p,k,l}\epsilon_{p}(k)c_{pkl}^{\dagger}c_{pkl} \label{hamilt}
\end{eqnarray}
In (\ref{hamilt}), TLS is describe by the pseudo-spin $\sigma$. The  two
eigenstates of $\sigma_{z}$,  $\mid\pm>$  correspond  to  the  tunneling
particle being in the  left  and  the  right  wells,  respectively.  The
electrons are described by a spherical wave representation, due  to  the
presence of a point scatterer in the Fermi sea. The first two  terms  in
(\ref{hamilt}) describe the bare TLS,  $\epsilon$  being  the  asymmetry
between the two wells and $\Delta_{0}$ the bare tunneling frequency. The
third term  represents  the  screening  of  the  TC  by  the  conduction
electrons. The fourth term in (\ref{hamilt}) describes electron assisted
tunneling, $\Delta_{1}$ being the strength of the process. The last term
is the Hamiltonian of the electrons in the partial wave  representation.
The indices $k$ and $k'$ represent the wave-vectors  of  the  electrons;
$l$ is the index for the partial wave and in the present model runs over
two values, $+$ and $-$; $p$ represents the spin of  the  electrons  and
runs over two values $\pm 1/2$; $\epsilon_{p}(k)$ is the energy  of  the
electrons with wave-vector $k$ and spin $p$; $\rho$ is  the  density  of
states at the Fermi level of the electron gas; $K$  is  a  dimensionless
coupling constant which describes the screening interaction of  the  TLS
with the electrons, whose value is restricted to be  $0\leq  K\leq  0.5$
\cite{Yamada}.   Operator   $c_{pk+}^{\dagger}$   ($c_{pk+}$)    creates
(annihilates) an electron in a state with spin $p$, a  wave-vector  $k$,
energy $\epsilon_{p}(k)$ and partial wave state +.  The  energy  of  the
excitations is assumed to have an  upper  cutoff,  equal  to  the  Fermi
energy  of  the  electron  gas,  $\epsilon_{F}$.  The   Hamiltonian   in
(\ref{hamilt}) can, infact,  be  thought  of  as  a  highly  anisotropic
two-channel  Kondo  Hamiltonian  with  two  external   magnetic   fields
$\epsilon$ and $\Delta_{0}$, pointing along z- and x-axis  respectively.
It also differs crucially from the Kondo problem due the fact  that  the
value of the so-called $J_{\parallel}$, in  the  present  case  is  very
different. Due to this, the present model shows coherent oscillation  of
the pseudo-spin, which is absent in in the conventional model.

        We have shown earlier that the dynamics of  the  single  channel
Kondo problem can be efficiently studied by bosonizing  the  Hamiltonian
\cite{tqkondo}. The success of this method in treating the effect of the
low-energy excitations on the impurity spin motivates us to  also  apply
it to the present problem. The two spin channels of the electrons  being
completely independent, the bosonization of one does not affect that  of
the other.  Consequently, the bosonized form for (\ref{hamilt})  can  be
written down in a manner similar to that of  the  single  channel  Kondo
Hamiltonian (for a  detailed  account  of  bosonization  the  reader  is
referred to \cite{Guinea}. The resulting Hamiltonian  then  assumes  the
following form

\begin{eqnarray}
H' &=& {1\over 2}\hbar\epsilon\sigma_{z} + {1\over 2}\hbar\Delta_{0}
\sigma_{x} \nonumber\\
&+& {1\over 2}\hbar\sqrt{2\pi K v_{F}/L}\sigma_{z}\sum_{k,p} \sqrt{\omega_{k}}
 e^{\omega_{k}/2\omega_{c}}(b_{pk}^{\dagger} + b_{pk}) \nonumber\\
&+& {1\over 2}\hbar\Delta_{1}\sum_{p}[\sigma_{+}e^{-\xi_{p}}
+ \sigma_{-}e^{+\xi_{p}}]
+ \sum_{kp} \hbar\omega_{k} b_{pk}^{\dagger}b_{pk}
\label{bozo}
\end{eqnarray}
where $\xi_{p}  =  {1\over  2}  \sum_{p}  (\pi  v_{F}/L\omega_{k})^{1/2}
\exp(-    \omega_{k}/2\omega_{c})     (b_{pk}^{\dagger}-b_{pk})$.     In
(\ref{bozo}), $p$ is the channel index, $\omega_{k} =  v_{F}k$,  $v_{F}$
being the Fermi velocity and $\omega_{c}$ is  a  high  frequency  cutoff
related to $\epsilon_{F}$. $L$ is the length of  the  normalization  box
and  the  limit  $L\rightarrow\infty$  is   taken   such   that   $(2\pi
/L)\sum_{k}\rightarrow \int dk$. The  bose  operator  $b_{pk}^{\dagger}$
creates a particle in the state $k$ of the $p$`th channel, which  is  an
excitation of the electron gas. The charge density excitations drop  out
of the problem.

  If $\Delta_{1}$  is  zero,  the  Hamiltonian  given  by  (\ref{bozo}),
without $\Delta_{0}$, can be diagonalized exactly. The strategy then, is
to treat the terms proportional to $\Delta_{0}$ and  $\Delta_{1}$  as  a
perturbation. To  this  end  we  perform  a  unitary  transformation  $U
H'U^{-1}$     on     the     Hamiltonian     where     $U\equiv     \exp
[\sigma_{z}\sqrt{K/2}\sum_{p} \xi_{p}]$. The transformed Hamiltonian has
the following form

\begin{eqnarray}
H &=& {1\over 2}\hbar\epsilon\sigma_{z} + \sum_{k,p} \hbar\omega_{k}
b_{pk}^{\dagger}b_{pk} \nonumber\\
&& + {1\over 2}\hbar\Delta_{0}[\sigma_{+}e^{\sqrt{K\over2}(\xi_{1} +\xi_{2})}
+\sigma_{-}e^{-\sqrt{K\over2}(\xi_{1} + \xi_{2})}]
+ {1\over 2}\hbar\Delta_{1}
 \nonumber\\
 &&\times\sum_{p}[\sigma_{+} e^{\sqrt{K\over2}(\xi_{1}+
 \xi_{2})-\xi_{p}} + \sigma_{-}e^{-\sqrt{K\over2}(\xi_{1}+\xi_{2}) +
 \xi_{p}}]
\label{trans}
\end{eqnarray}
One can see that for $\Delta_{0}$, $\Delta_{1} = 0$, the Hamiltonian  in
(\ref{trans}) is diagonal in the usual representation. In the  following
we denote the first and the second terms of (\ref{trans}) by $H_{S}$ and
$H_{B}$, respectively, and the other two terms by  $H_{I}$.  A  suitable
perturbation  theory  can  now  be  done  on  $H$  by  treating  $H_{I}$
perturbatively, which amounts to  treating  $K$  exactly  and  doing  an
expansion in $\Delta_{0}$ and $\Delta_{1}$.

        The quantity of interest here is the Laplace  transform  of  the
time correlation function of the  pseudo-spin,  $C(t)  =  <\sigma_{z}(0)
\sigma_{z}(t)>$, given by

\begin{equation}
\hat{C}(z) = \int^{\infty}_{0} e^{-zt} <\sigma_{z}(0)\sigma_{z}(t)> dt.
 \label{czdef}
\end{equation}
Here, the angular brackets denote canonical ensemble  average,  and  the
Heisenberg time evolution of $\sigma_{z}$ is dictated  by  $H$.  In  the
following analysis, the Greek indices $\mu$, $\nu$ denote  the  impurity
spin states, and the ``states'' of the Liouville operators  are  denoted
by $\mid\nu\mu)$ etc. We shall now focus our attention  on  $\hat{C}(z)$
which can be put in the following form:

\begin{eqnarray}
C(z)&=&\sum_{\nu,\nu'}<\nu\mid\rho_{S}\mid\nu><\nu\mid
S_{z}\mid\nu>\nonumber\\
&&(\nu,\nu\mid [U( z)]_{av} \mid\nu',\nu')<\nu'\mid S_{z}\mid\nu'> \label{cz}
\end{eqnarray}
In writing (\ref{cz})  we  have  made  use  of  the  Liouville  operator
formalism to  introduce  a  ``bath-averaged  time  evolution  operator''
$[U(t)]_{av}$,  where  $U(t)  =  e^{iLt}$,  $L$  being  the  Liouvillian
associated  with  $H$.  In  addition, we have factorized  the  canonical
density matrix $\rho$, as $\rho \approx \rho_{S}\cdot  \rho_{B}$,  where
$\rho_{S}$  is  the  density  operator  associated  with   $H_{S}$   and
$\rho_{B}$ is the density operator for a bath of  noninteracting  bosons
given by $H_{B}$. The bath average $[...]_{av}$ implies a multiplication
by $\rho_{B}$ and a trace over the bath states. The  bath-averaged  time
evolution operator contains all the information regarding the  influence
of   the   electronic   environment   on   the   TLS.    We    calculate
$[\hat{U}(z)]_{av}$ using the resolvent expansion  formalism  where  the
resolvent is treated perturbatively to yield a ``self-energy'' which  is
second  order  in  $H_{I}$,  resulting  in  $[\hat{U}(z)]_{av}   \approx
[z-iL_{S}  +   \{L_{I}(z-iL_{S}-iL_{B})^{-1}L_{I}\}_{av}]^{-1}$,   where
$L_{S}$, $L_{I}$ and $L_{B}$ are  Liouville  operators  associated  with
$H_{S}$,  $H_{I}$   and   $H_{B}$   respectively   \cite{Grabert}.   The
calculation proceeds along the lines of Ref. \cite{Grabert} which treats
the dynamics of a TLS in contact with a bosonic bath.

We first calculate the matrix for the self-energy $[L_{I} (z - iL_{S}  -
iL_{B})^{-1} L_{I}]_{av}$ and it turns out  that  in  the  second  order
resolvent  expansion,  it  can  be  represented  in  terms  of   certain
correlation functions  of  a  gas  of  free  bosons.  Infact,  one needs
quantities like  $<e^{+a\xi_{1}}\cdot e^{-b\xi_{1}(t)}>_{B}$  where  the
subscript $B$ indicates that  the  correlation  function  is  calculated
using only a free-boson Hamiltonian  \cite{Grabert}.  These  correlation
functions can be calculated using certain well known properties of a set
of   harmonic   oscillators   using    the    approximation    $T    \ll
\hbar\omega_{c}/k_{B}$ and  $t  \ll  1/\omega_{c}$  \cite{Grabert}.  The
self-energy, being related to the Liouville operators, is  a  4$\times$4
matrix within the space of the  states  of  the  pseudo-spin.  But  this
matrix turns out to be block diagonal, and the $2\times 2$ blocks can be
handled with ease.  Consequently,  the  block-diagonal  matrix  for  the
self-energy is combined with the diagonal matrix for $(z - iL_{S})$  and
inverted to yield the averaged time-evolution operator $[\hat{U}(z)]_{av}$.

\section{Result}

Once the bath-averaged time-evolution operator is known, we are all  set
to calculate the Laplace transformed correlation function, the  quantity
of central importance to us. We plug in the form of  $[\hat{U}(z)]_{av}$
in (\ref{cz}) to yield :

\begin{equation}
\hat{C}(z) = {z+\tanh{(\hbar\epsilon\beta/2)} [\Phi_{-}(z+i\epsilon) -
\Phi_{-}(z-i\epsilon)] \over z [ z + \Phi_{+}(z+i\epsilon) +
\Phi_{+}(z-i\epsilon)] }, \label{result}
\end{equation}
where $\Phi_{\pm}(z) = F(z) \pm F'(z)$, and

\begin{eqnarray}
F(z) &=& {1\over 4} \Delta_{0}^{2} (\Theta/\omega_{c})^{2K} {\Gamma(1-2K)
\Gamma(1+K+z/\Theta) \over (z + \Theta K) \Gamma(1-K+z/\Theta)} e^{i\pi K}
\nonumber\\
&-& {\Delta_{1}^{2}\over 2\omega_{c}^{2}}
{\Gamma(2\eta-1) \Gamma(1-\eta+z/\Theta) \over
(\Theta/\omega_{c})^{2\eta}\Gamma(1+\eta+z/\Theta)} (z+\eta\Theta)
 e^{-i\pi\eta} \nonumber\\
&+& {\Delta_{1}^{2}\over 32\pi^{4}} {\Gamma(2\eta-1)
\Gamma(1-\eta+z/\Theta) \over (z - \eta\Theta)
\Gamma(1+\eta+z/\Theta)} \left(\Theta\over\omega_{c}\right)^{4-2\eta}
e^{i\pi \eta} \nonumber\\
&+& {\Delta_{0}\Delta_{1}\over 4\pi^{2}} {\Gamma(1+2\delta) \Gamma(1-
\delta+z/\Theta) \over (z - \delta\Theta) \Gamma(1+\delta+z/\Theta)}
\left(\Theta\over\omega_{c}\right)^{2-2\delta} e^{i\pi \delta}. \nonumber\\
\label{Fz}
\end{eqnarray}
In  (\ref{Fz}),   $\Theta\equiv   2\pi   k_{B}T/\hbar$,   $\eta   \equiv
\sqrt{2K}-K$, $\delta \equiv \sqrt{2K} - 2K$ and $F'(z)$ is the same  as
$F(z)$ except that the exponential factors are replaced by their complex
conjugates. Some comments at this stage. The expression for the  Laplace
transformed correlation function, given by (\ref{result}), describes the
dynamics of a TC in a metal. The dynamics is governed by  the  screening
of the TC by the conduction electrons, as well as the electron  assisted
process. The first term represents the effect of screening of the TC  by
the conduction  electrons,  while  the  second  term  accounts  for  the
electron assisted process. The third and the fourth terms in  (\ref{Fz})
are much smaller in magnitude compared to the first two, because of  the
presence of the extra factors $\Theta/\omega_{c}$.

\section{Discussion and conclusion}

The  second  order  treatment  of  $\Delta_{0}$  and   $\Delta_{1}$   is
equivalent to some kind of a DBGA if one were to employ  the  functional
integral formalism, pioneered by Leggett {\it et al} \cite{Leggett}. Let
us, for the time being, neglect electron assisted tunneling, by  putting
$\Delta_{1}=0$, which reduces the expression in  (\ref{result})  to  the
following

\begin{equation}
\hat{C}(z) = {z+i\tanh{({1\over 2}\hbar\epsilon\beta)}\sin{\pi K}
[G(z+i\epsilon) - G(z-i\epsilon)] \over z [ z + \cos{\pi K}\{G(z+i\epsilon)
+ G(z-i\epsilon)\}] }, \label{dbga}
\end{equation}
where

\begin{equation}
G(z) = {1\over 2} \Delta_{0}^{2} (\Theta/\omega_{c})^{2K} {\Gamma(1-2K)
\Gamma(1+K+z/\Theta) \over (z + \Theta K) \Gamma(1-K+z/\Theta)}.
\end{equation}
This is just the DBGA result for a dissipative asymmetric TLS, for ohmic
dissipation, derived by Leggett {\it et  al}  \cite{Leggett}  using  the
path-integral  formalism, and by Dattagupta {\it et  al}  \cite{Grabert}
using   the   resolvent   expansion   formalism.   By   including    the
$\Delta_{1}$-dependent terms  in  the  expression  for  the  correlation
function, we extend the DBGA result  to  the  case  where  the  electron
assisted process plays an important role, in addition to  the  screening
effect.

In order to make explicit analysis simpler we  look  at  the  particular
case of a symmetric double-well where $\epsilon=0$.  The  real  part  of
$\hat{C}(z)$,  which  may  describe  the  neutron  scattering  structure
factor, is in general not a Lorentzian. In order to force  a  Lorentzian
form, we further assume that $K\ll 1$.  With  this  simplification,  the
expression (\ref{result}) can be approximated by

\begin{equation}
\hat{C}(z) \approx {1\over z+\tilde{\Delta}_{0}^{2}/(z+\gamma_{0})+
2\tilde{\Delta}_{1}^{2}
(z+\gamma_{1}) },
\label{symmetric}
\end{equation}
where    $\tilde{\Delta}_{0}     =     \Delta_{0}(\Theta/\omega_{c})^{K}
[\Gamma(1-2K)     Cos(\pi     K)]^{1/2}$,     $\tilde{\Delta}_{1}      =
(\Delta_{1}/\omega_{c})   (\omega_{c}/   \Theta)^{\eta}   [\Gamma(2\eta)
Cos(\pi \eta)/(1-2\eta)]^{1/2}$, $\gamma_{0} = K\Theta$ and  $\gamma_{1}
= \eta\Theta$. In writing (\ref{symmetric}) we have neglected  the  last
two    terms     in     (\ref{Fz})     (recall     the     approximation
$\omega_{c}\gg\Theta$).  The  real  part  of  the  Laplace   transformed
correlation function given by (\ref{symmetric}) describes two Lorentzian
lines   of   width   $\gamma_{L} = \gamma_{0} +  2\tilde{\Delta}_{1}^{2}
\gamma_{1}/ (1 + 2\tilde{\Delta}_{1}^{2})$,  centered  at  $\omega
=\pm[\tilde{\Delta}_{0}^{2}/(1+2\tilde{\Delta}_{1}^{2})-{\gamma_{0}+
2\tilde{\Delta}_{1}^{2}\gamma_{1}/(1+2\tilde{\Delta}_{1}^{2})}]^{2}$.
One can see that the tunnel frequency of the  particle  is  renormalized
due to the effect of metallic electrons. The renormalization is not only
due to the overlap of electron states corresponding to the two positions
of the  particle  (which  leads  to  a  reduction  factor  of  the  form
$(k_{B}T/\hbar\omega_{c})^{K})$ but also due to the effect  of  electron
assisted process. This will show up in the effective tunnel splitting of
a particle, at very  low  temperatures  where  it  is  in  the  coherent
tunneling  regime.  The   width   of   the   peaks,   because   of   the
$\tilde{\Delta}_{1}$  dependent  term,  also  deviates  from  the  usual
``Korringa form'' at very low temperatures.

        The DBGA is known to break down at  very  low  temperatures  for
small frequencies. In order  to  circumvent  this  difficulty  we  shall
concentrate on the  real  part  of  $\hat{C}(z)$  near  the  resonances.
Without any interaction with electrons the resonances for the TLS are at
$z=\pm i\Delta_{0}$. So, we replace $F(z)$ in the expression  (\ref{Fz})
for $\hat{C}(z)$, by $F(i\Delta_{0})$.  Thus,  the  relaxation  rate  is
given by $\gamma = Re[\hat{C}(i\Delta_{0})]$, and  the  tunnel-splitting
is given by $\omega = -Im[\hat{C}(i\Delta_{0})$.  Figures  \  \ref{rate}
and \ref{tunnel} show plots of the two against temperature. In the plots,
everything  is  scaled  with  $\Delta_{0}$,   to   yield   dimensionless
variables. One can see that for low  enough  temperature,  the  electron
assisted process has a dramatic effect on the  relaxation  rate. At very
low  temperature,  such  a system  is known  to  go  into  a  Kondo-like
correlated state. This may be the reason for this increased contribution
to relaxation. This feature can possibly be experimentally observed. The
tunnel-splitting, however,  does  not  show  a  significant  qualitative
change in its  temperature  dependence.  The plots indicate that at high
temperatures the electron-assisted process does not have  a  significant
effect. But as the temperature  is  lowered, it  is  this process  which
dominates the dynamics of the TLS.

  In conclusion, we have  studied  the  dynamics  of  a  particle  in  a
double-well potential, in the presence of conduction electrons where the
effect of electron assisted tunneling is taken into account in  addition
to conventional damping effects. In the appropriate limit  the  relevant
expression reduces to the DBGA result  for  the  dynamics  of  an  Ohmic
dissipative two-level system. In  the  coherent  tunneling  regime,  the
tunnel frequency is modified by the effect of electrons. The  relaxation
rate shows an unexpected rise as the temperature is lowered, as  opposed
to the case where the electron assisted process is  absent.  We  believe
that this feature should be observable in a careful experimental study.

\begin{acknowledgements}
Part of this work was completed during the author's stay at  the  School
of Physical Sciences, Jawaharlal Nehru  University,  and  he  wishes  to
thank people who made the stay possible. The author is thankful  to  the
Indian Institute of Technology, New Delhi, India for providing necessary
library facilities.

\end{acknowledgements}

\begin{figure}
\caption{For plotting the relaxation rate and tunnel  splitting  against
$\Theta$, we scale all quantities with $\Delta_{0}$, so that  $\Gamma  =
\gamma/\Delta_{0},\    \tau    =    \Theta/\Delta_{0},\     \Delta     =
\Delta_{1}/\Delta_{0}$ and $D = \omega_{c}/\Delta_{0}$. The  plot  shows
dimensionless relxation rate per temperature $\Gamma/\tau$  against  the
temperature $\tau$,\  for  $D=100.0$,\  $K=0.05$,\  $\Delta=1.0$  (solid
line), and $\Delta=0.0$ (dashed line). At low temperatures the  behavior
of the  solid  line  indicates  the  increased  relaxation  due  to  the
electron-assisted process.}\label{rate}
\end{figure}

\begin{figure}
\caption{$\Omega=-Im[\hat{C}(i\Delta_{0})/\Delta_{0}]$ plotted against
temperature, for $D=100.0$,\ $K=0.05$,\ $\Delta=1.0$ (solid line), and
$\Delta=0.0$
(dashed line).}\label{tunnel}
\end{figure}
\end{narrowtext}

\begin{references}
\bibitem[\dagger]{email} Email : tabish@imsc.ernet.in
\bibitem{Black} J.L. Black, in {\it Glassy Metals I}, edited by H.J.
Guntherodt
and H. Beck (Springer-Verlag, Berlin, 1981), p. 167
\bibitem{Wipf} H. Wipf and K. Neumaier, Phys. Rev. Lett. 52, 1308 (1984)
\bibitem{Grabert} S. Dattagupta, H. Grabert and R. Jung, J. Phys. : Cond.
Matt. 1, 1405 (1989)
\bibitem{Weiss} U. Weiss and Wollensak, Phys. Rev. Lett. (1989)
\bibitem{tqniobium} S. Dattagupta and T. Qureshi, Physica B 174, 262 (1991)
\bibitem{Ralph} D.C. Ralph, A.W.W. Ludwig, J. von Delft and R.A. Buhrman,
 Phys. Rev. Lett. 72, 1064 (1994)
\bibitem{Ishiguro} T. Ishiguro, H. Kaneko, Y. Nogami, H. Ishimoto, H.
Nishiyama,
 J. Tsukamoto, A. Takahashi, M. Yamamura, T. Hagiwara and K. Sato, Phys.
  Rev. Lett. 69, 660 (1992)
\bibitem{Golding} B. Golding, N.M. Zimmerman ans S.N. Coppersmith, Phys. Rev.
Lett. 68, 998 (1992)
\bibitem{Leggett} A.J. Leggett, S. Chakravarty, M.P.A. Fisher, A.T. Dorsey,
A. Garg and W. Zwerger, Rev. Mod. Phys. 59, 1 (1987)
\bibitem{Kondo} J. Kondo, Physica (Amsterdam) 84B, 207 (1976)
\bibitem{Zawa} A. Zawadowsky, Phys. Rev. Lett. 45, 211 (1980)
\bibitem{Muramatsu} A. Muramatsu and F. Guinea, Phys. Rev. Lett. 57, 2337
(1986)
\bibitem{Vladar} K. Vlad\'{a}r and A. Zawadowsky, Phys. Rev B 28, 1564 (1983);
28, 1582 (1983)
\bibitem{Zimanyi}  K. Vlad\'{a}r, G. Zim\'{a}nyi and A. Zawadowsky, Phys. Rev.
Lett. 56, 286 (1986)
\bibitem{Zarand} G. Zar\'{a}nd and A. Zawadowsky, Phys. Rev. Lett. 72, 542
 (1994)
\bibitem{Nozieres} P. Nozi\`{e}res and A. Blandin, J. de Phys. 41, 193 (1983)
\bibitem{pramana} T. Qureshi and S. Dattagupta, Pramana-J. Phys. 35, 579
(1990)
\bibitem{Ludwig} A.W.W. Ludwig and I. Affleck, Phys. Rev. Lett. 25, 3160
(1991)
\bibitem{Yamada} K. Yamada, A. Sakurai and S. Miyazima, Progress Theor. Phys.
73, 1342 (1985)
\bibitem{tqkondo} T. Qureshi and S. Dattagupta, Phys. Rev. B 49, 12848 (1994)
\bibitem{Guinea} F. Guinea, V. Hakim and A. Muramatsu, Phys. Rev. B 32, 4410
(1985); K.D. Schotte, Z. Phys. 230, 99 (1970); P. Schlottmann, Phys. Rev. B
  25, 4815 (1982)

\end{references}
\end{document}